\documentclass[prl,aps,floats,twocolumn,superscriptaddress,floatfix,longbibliography]{revtex4-1}
\usepackage{amssymb,amsmath}
\usepackage{amsmath,amssymb}
\usepackage[mathscr]{euscript}
\usepackage{graphicx}
\usepackage{psfrag}
\usepackage[normalem]{ulem}
\usepackage{xcolor}
\usepackage[%
  colorlinks=true,
  urlcolor=blue,
  linkcolor=blue,
  citecolor=blue
]{hyperref}

\def\beq{\begin{equation}}
\def\eeq{\end{equation}}
\def\bea{\begin{eqnarray}}
\def\eea{\end{eqnarray}}

\definecolor{mygreen}{rgb}{0.0,0.55,0.3}

\newcommand{\rlj}[1]{{\color{mygreen}#1}}

\newcommand{\rhobar}{\bar\rho}
\begin{document}
 \title{Role of initial conditions in $1D$ diffusive systems: compressibility, hyperuniformity and long-term memory}
\author{Tirthankar Banerjee}\email{tb698@cam.ac.uk}
\affiliation{DAMTP, Centre for Mathematical Sciences, University of Cambridge, Wilberforce Road, Cambridge, CB3 0WA}
\author{Robert L. Jack}
\affiliation{DAMTP, Centre for Mathematical Sciences, University of Cambridge, Wilberforce Road, Cambridge, CB3 0WA}
\affiliation{Yusuf Hamied Department of Chemistry, University of Cambridge, Lensfield Road, Cambridge CB2 1EW, United Kingdom}
\author{Michael E. Cates}
\affiliation{DAMTP, Centre for Mathematical Sciences, University of Cambridge, Wilberforce Road, Cambridge, CB3 0WA}

\date{\today}
\begin{abstract}
We analyse the {long-lasting} effects of initial conditions on
dynamical fluctuations in one-dimensional diffusive systems.  We consider the mean-squared
displacement of tracers in homogeneous systems with single-file
diffusion, and current fluctuations  for non-interacting diffusive particles. 
In each case we show analytically that the long-term
memory of initial conditions is mediated by a single static quantity: a
generalized compressibility that quantifies the density fluctuations of
the initial state. We thereby identify a universality class of
hyperuniform initial states {whose dynamical variances coincide with the
`quenched' cases studied previously, alongside a continuous
family} of other classes among which equilibrated (or `annealed') initial
conditions are but one member. We verify our {predictions} through extensive Monte Carlo simulations.
\end{abstract} 

\maketitle

In single-file diffusion, particles move in a single lane with no possibility of overtaking~\cite{Bechinger-Science,Hodgkin, harris,richards,pincus,arratia,sadhu-prl,hegde-prl,barkai-pre, kundu-cividini, single-file-review}. Modern technology has made this process increasingly relevant in experimental~\cite{vgupta-chemphyslett, kukla-science, sood-acs-nano, kollman-bechinger, rasaiah-ann-rev}, industrial~\cite{zeolite, karger-zeolite, single-file-review}, biological~\cite{nanopore-dna} and biomedical~\cite{protein-nano} settings. 
Of fundamental interest is the mean-squared displacement (MSD) 
of a tagged particle (tracer) within a collection of identical particles executing single-file diffusion.  This MSD grows with time as $\sqrt{t}$, in contrast to normal diffusion, where it grows as $t$.  

Remarkably, theoretical studies~\cite{satya-mustansir, van-beijeren, Rajesh-Satya, barkai-pre, sadhu-prl, kundu-cividini, rana-sadhu-arxiv} have shown that MSDs can depend on the initialization of the system, prior to measurement.   For example, in 1D one might either initialise point particles uniformly at random, or prepare an initial state with equi-spaced particles.  The MSD grows as $\sqrt{t}$ in both cases, but the prefactors are different~\cite{barkai-pre}. This is an {\em everlasting} dependence on the initial state, affecting the asymptotic behavior at large times.
Similar everlasting effects of the initialization protocol are also observed in other 1D systems (not necessarily single-file), for example when measuring the variance of particle currents~\cite{derr-gers, tirtha-paris}.  
In both cases, one may additionally choose to perform either a quenched or annealed average over initial states~\cite{derr-gers,sadhu-prl}: this choice also has an everlasting effect on the resulting behavior.
These results establish that 1D particle systems can retain long-term memory of their initialization: this is a form of non-ergodicity, which may occur (in general) by many different mechanisms~\cite{Bouchaud-92, Bel-Barkai,schutz-rapidcomm-pre,boudini,Palmer-review,Goldenfeld-book,FPU,FPU-review,derr-gers,metzler-njp2013}.

We show here that these systems' long-term memory can be explained by a unified framework, greatly generalizing previous results known for specific cases~\cite{kollmann-prl,derr-gers,barkai-pre,tirtha-paris,sadhu-prl,Lizana-pre, illien-2022}.   Physically, we note that the large-scale density fluctuations of the initial condition relax very slowly: this causes long-term memory.  
Remarkably, the role of these fluctuations can be quantified by a single static quantity -- the Fano factor~\cite{Fano} or generalized compressibility:
\begin{equation}\label{fano}
\alpha_{\rm ic} = \lim_{\ell \to\infty} \frac{ {\rm Var}[n(\ell)] }{ \overline{n}(\ell) } \, .
\end{equation}
Here $n(\ell)$ is the number of particles initially found within a distance $\ell$ of the origin and $\overline{n}(\ell)$ is its average; $\alpha_{\rm ic}$ encapsulates the effects of initial density fluctuations on systems' asymptotic (large-time) dynamical behavior.
Notably, $\alpha_{\rm ic}$ in \eqref{fano} is a static property of the initial state that is experimentally accessible ({\em e.g.,} via microscopy) without knowing details of particle interactions~\cite{alpha-2d-exp, alpha-classes-exp, Keyser-current}.  Via Eq.~(\ref{variance-tracer-intro}) below, it directly controls a dynamical quantity that is similarly accessible~\cite{kollman-bechinger}, namely the variance of particle displacements up to time $T$.

We now summarize our main results, before presenting their derivation. 
First, consider single-file diffusion 
 of a homogeneous interacting particle system under conditions where Macroscopic Fluctuation Theory (MFT) is applicable~\cite{sadhu-prl, MFT-main,derrida-mft,bodineau-prl,mft-group-prl,jona-lasinio-jstat, mallick-mft-sep-bethe, mallick-mft-sep-exact}. MFT is a hydrodynamic theory that captures numerous microscopic processes~\cite{MFT-main} including the simple exclusion process, point Brownian particles, zero range processes, 
 and random average processes~\cite{Rajesh-Satya, illien-prl}. 
The system is initialized in a macroscopically homogeneous state with mean density $\rhobar$ and density fluctuations determining $\alpha_{\rm ic}$ via \eqref{fano}.
 Identifying a single tracer particle starting from the origin, we will show that the variance of its position $X(T)$ 
 satisfies
\begin{equation}\label{variance-tracer-intro}
\operatorname{Var}[X(T)] \simeq  \Delta X^2_{\rm noise}(T{,\rhobar}) + \alpha_{\rm ic} \Delta X^2_{\rm dens}(T{,\rhobar}),
\end{equation}
where  `$\simeq$' indicates asymptotic equality for large $T$. Expressions for $\Delta X^2_{\rm noise}$ and $ \Delta X^2_{\rm dens}$ are given in \eqref{deltax2} and \eqref{deltax1}, respectively. Both are proportional to $\sqrt{T}$, and depend on $\rhobar$ and on the transport coefficients -- diffusivity $D(\rhobar)$ and mobility $\sigma(\rhobar)$ encoded in MFT~\cite{MFT-main,derrida-mft}.  The variance in \eqref{variance-tracer-intro} includes the stochastic motion of the particles in the system, and any random aspects of the initialization process; it corresponds to an `annealed' variance in the terminology of~\cite{derr-gers}.

Several previous results are special cases of Eq.~\eqref{variance-tracer-intro}.
If the initial condition is the {thermal} equilibrium ensemble of the system's own dynamics, then
$\alpha_{\rm ic}$ is the thermal compressibility factor~\cite{chaikin-lubensky, bell-fcthm},
and we recover a result of~\cite{kollmann-prl,sadhu-prl}.  
If in contrast one chooses initial conditions that are hyperuniform (HU), then $\alpha_{\rm ic}=0$ by definition~\cite{torquato-2018} and \eqref{variance-tracer-intro} correctly predicts a different variance, $\Delta X^2_{\rm noise}$.
This was known for the equi-spaced initial condition of~\cite{barkai-pre}, which now emerges as representative of a much larger, HU universality class. Another special case was obtained in~\cite{sadhu-prl} by considering a quenched average over initial conditions: in our framework, this also represents $\alpha_{\rm ic}=0$ (see~\cite{SM}).
Moving beyond these special cases, we emphasize that \eqref{variance-tracer-intro} reveals a continuous spectrum of classes with variances $\operatorname{Var}[X(T)] $ parameterised by $\alpha_{\rm ic}$, which is tunable via the initialization protocol.
 
To elucidate the physics behind results (\ref{fano},\ref{variance-tracer-intro}), we also analyze below a diffusive system of {\em non-interacting} particles.  After initialising a homogeneous system at density $\rhobar$, we remove all particles to the right of an arbitrary origin.  The remaining particles have dynamics that at large times is Brownian with diffusivity $D$.  Let $Q(T)$ be the integrated flux of particles through the origin, up to time $T$.  
We shall show that
\begin{equation}\label{variance-nonint-intro}
\operatorname{Var}[Q(T)] \simeq \Delta Q^2_{\rm noise}(T{,\rhobar}) + \alpha_{\rm ic} \Delta Q^2_{\rm dens}(T{,\rhobar}) \, .
\end{equation}
Here $\Delta Q^2_{\rm noise}$
 and $\Delta Q^2_{\rm dens}$, given explicitly in \eqref{V1-fin} and \eqref{V2-fin}, respectively, both grow as $\sqrt{T}$ for large times, and $\alpha_{\rm ic}$ again obeys \eqref{fano}.

The similarity between (\ref{variance-tracer-intro},\ref{variance-nonint-intro}) is striking: 
both variances depend on initialization solely via the mean density $\rhobar$ 
and the Fano factor $\alpha_{\rm ic}$.  
Moreover, they share a physical origin: everlasting memory of the initial condition can only survive through modes whose lifetime is unbounded. These are the (conserved) mean density, and the large-scale ($\ell\to\infty$) density fluctuations, quantified by $\alpha_{\rm ic}$.

We next derive the above results, first obtaining \eqref{variance-nonint-intro} by direct computation.
To derive \eqref{variance-tracer-intro} is more complicated, but the same physical principles are at work. 
Indeed, while \eqref{variance-tracer-intro} is more relevant for applications involving single-file diffusion, \eqref{variance-nonint-intro} provides a simpler illustration of the underlying physics.

{\em Current fluctuations for non-interacting particles:}  Consider non-interacting particles with initial positions $\textbf{y}=(y_1,y_2,\dots)$ {lying} to the left of the origin ($y_i<0$)  At time $t$, define $\chi_i(t)=1$ if particle $i$ has position $x_i(t)>0$, and $\chi_i(t)=0$ otherwise.  
Also define the propagator $G(x,y,t)$ as the probability density that a particle is at position $x$, given that it was at position $y$ a time $t$ earlier.  For {given initial conditions} $\textbf{y}$, {we have} $\langle \chi_i(t) \rangle_\textbf{y} = U(-y_i,t)$,
where~\cite{tirtha-paris}
\begin{equation}
U(z,t) = \int_0^{\infty} \!dx\, G(x,-z,t) \,.
\end{equation}
The notation $\langle ... \rangle_{\textbf{y}}$ represents an average over the stochastic particle dynamics, for a given initial condition $\textbf{y}$.
{The mean integrated flux} through the origin follows by summing over all particles:
 $\langle Q(t)\rangle_{\textbf{y}} = \sum_i U(-y_i,t)$. 
Defining the empirical density of the initial condition $\hat{\rho}(x|\textbf{y}) = \sum_i  \delta(x-y_i)$, we write
\begin{eqnarray}\label{meanflux-rho}
\langle Q(t)\rangle_{\textbf{y}} =\int_{0}^{\infty} \!dz\, \hat{\rho}(-z|\textbf{y}) U(z,t) \; .
\end{eqnarray}
(Here and below, positive $z$ lies to the \emph{left} of the origin.)

For independent particles, and since {$\chi_i\in \{0,1 \}$}, the quantity
\begin{equation}\label{dyn-Qvar}
\langle Q(t)^2 \rangle_{\textbf{y}} 
-\langle Q(t) \rangle^2_{\textbf{y}} = \int_{0}^{\infty} \!dz\, \hat{\rho}(-z|\textbf{y})
U(z,t)[1- U(z,t)] .
\end{equation}
measures how much $Q(t)$ fluctuates between trajectories, for a fixed initial condition $\textbf{y}$.  
For many initialization protocols, the initial condition is itself stochastic, so
the next step is 
to average over $\textbf{y}$, (denoted as $\overline{(...)}$).  We define the variance of the 
flux to include both sources of randomness~\cite{SM,ludovic-rob}:
\begin{equation}\label{VQ-V1V2}
\operatorname{Var}[Q(t)] =  \Delta Q^2_{\rm noise}(t{,\rhobar})  + \Delta Q^2_{\rm ic}(t{,\rhobar}) 
\end{equation}
where 
\begin{align}
 \Delta Q^2_{\rm noise} & = \overline{\langle Q(t)^2 \rangle_{\textbf{y}} } -  \overline{ \langle Q(t) \rangle_{\textbf{y}}^2} \; ,
\nonumber\\
\Delta Q^2_{\rm ic}(t)  &=  \overline{ \langle Q(t) \rangle_{\textbf{y}}^2}  - \overline{ \langle Q(t) \rangle_{\textbf{y}}}^2 \; .
\end{align}
In the disordered-systems terminology of~\cite{derr-gers}, $\operatorname{Var}[Q(t)]$ and $ \Delta Q^2_{\rm noise}$ are respectively ``quenched'' and ``annealed" variances~\cite{Palmer-review,Goldenfeld-book,Binder-review}.  Physically, $\Delta Q^2_{\rm noise}$ measures how much $Q$ fluctuates between trajectories with the same initial condition, while $\Delta Q^2_{\rm ic}$ depends additionally on the fluctuations of the initial condition, which are never forgotten.
Using \eqref{meanflux-rho} yields
\begin{multline}\label{4V2-1}
\Delta Q^2_{\rm ic}(t)= \int_{0}^{\infty} \int_{0}^{\infty}\!dz \, dz'\,  
U(z,t)U(z',t)  C_2(z,z')
\end{multline}
where {$C_2(z,z') = \overline{\hat{\rho}(-z|{\bf y}) \hat{\rho}(-z'|{\bf y})} - \overline{\hat{\rho}(-z|{\bf y})} \, \overline{\hat{\rho}(-z'|{\bf y})}$}.  Hence, the  initial fluctuations enter the variance through the one- and two-point density correlations only.

So far, this analysis is general. We now specialize to the case where, for large $t$, the propagator $G$ is diffusive, such that 
\begin{equation}\label{Uzt-Brownian}
U(z,t) \simeq \frac{1}{2} {\rm erfc}\left[\frac{z}{\sqrt{4Dt}} \right] \quad \hbox{as $t\to\infty$} \; .
\end{equation}
This assumption covers passive diffusers, and many kinds of active particle whose late-time motion is also diffusive.
Second, we consider initial conditions found by taking an infinite, translationally invariant system and erasing all particles to the right of the origin. This means that {$ \overline{\hat{\rho}(-z|{\bf y})} = \rhobar \Theta(z)$} and $C_2(z,z') = \rhobar  \Theta(z) \Theta(z')C(z-z')$ where $\Theta(z)$ is the Heaviside function and $C(z-z')$ the two-point {correlator before erasure}.
These assumptions can be relaxed, but are sufficient here.

For large {times $T$}, 
 (\ref{4V2-1},\ref{Uzt-Brownian}) yield
\begin{multline}\label{V2S}
\Delta Q^2_{\rm ic}(T)
 \simeq  \frac{\rhobar\sqrt{D T}}{4 \pi} \int_{0}^{\infty} \!dy \int_{0}^{\infty} \!dy' \int_{-\infty}^{\infty}  \!dp \,
 e^{i p (y'-y)}   \\ \times  S\left(\frac{p}{\sqrt{4 D T}}\right) \, \, {\rm erfc} (y) \, {\rm erfc} (y')\, \,,
\end{multline}
where $S(q) = \int_{-\infty}^{\infty}  \!dz\, C(z) e^{-iqz}$ is the structure factor.
{The fluctuations of the initial condition enter} this expression solely through  $\alpha_{\rm ic} = \lim_{q\to0} S(q)$, which  is equivalent to (\ref{fano})~\cite{bell-fcthm, Landau-Lifshitz-Statmech, SM}.
Replacing the structure factor by its limit, the integrals in (\ref{V2S}) {yield}
$\Delta Q^2_{\rm ic}(T{,\rhobar}) =  \alpha_{\rm ic} \Delta Q^2_{\rm dens}(T{,\rhobar}) $ with 
\begin{equation}\label{V2-fin}
\Delta Q^2_{\rm dens}(T{,\rhobar})  {\,\simeq\,}  \big(\sqrt{2}-1\big) \sqrt{\frac{\rhobar^2 D T }{2 \pi}}\, \, .
\end{equation}
Similarly using \eqref{dyn-Qvar}, $ \Delta Q^2_{\rm noise}$ {is:}
\begin{equation}\label{V1-fin}
 \Delta Q^2_{\rm noise}(T{,\rhobar}) { \,\simeq\, } \sqrt{\frac{\rhobar^2 D T }{2 \pi}} \; .
\end{equation} 
Combining (\ref{VQ-V1V2},\ref{V2-fin},\ref{V1-fin}) gives the promised result,  Eq.~\eqref{variance-nonint-intro}.

These results confirm that the current variance has an everlasting dependence on the fluctuations of the initial state, through $\alpha_{\rm ic}$.  The case $\alpha_{\rm ic}=0$ arises if the initialization has no randomness at all ({\em e.g.}, equi-spaced particles) or is hyperuniform; the variance in these cases is simply
$ \Delta Q^2_{\rm noise}$, referred to as the `quenched' variance in~\cite{tirtha-paris,derr-gers} (see \cite{SM}).
On the other hand, if the initial condition has $C(z)=\delta(z)$ (as holds for an equilibrated ideal gas), then $\alpha_{\rm ic}=1$.
This coincides with the `annealed' variance computed in~\cite{tirtha-paris,derr-gers}, which is
larger than the quenched variance by a factor $\sqrt{2}$.
These previously-studied cases now emerge as two specific choices within an infinite family of classes of initial condition, {parameterized by} $\alpha_{\rm ic}$ which may take any non-negative value. 

To understand the physical mechanism, note that
together, (\ref{meanflux-rho},\ref{Uzt-Brownian}) imply that particles {starting {within} $\sqrt{4DT}$ from the origin have passed it with probability $1/2$} {after time $T$}, while particles starting {much further away} are unlikely to have done so.  Hence, the average {flux is} controlled by the number of particles within $\sqrt{4DT}$ of the origin. For large $T$ the variance of this number is determined by $\alpha_{\rm ic}$ via (\ref{fano}), thereby controlling $\Delta Q^2_{\rm ic}(T,\rhobar)$.  The everlasting effect of the initial conditions stems from the longest-wavelength density fluctuations that determine $\alpha_{\rm ic}$, {whose unbounded relaxation times are the} source of long-term memory.
\begin{figure}
\includegraphics[width=\columnwidth]{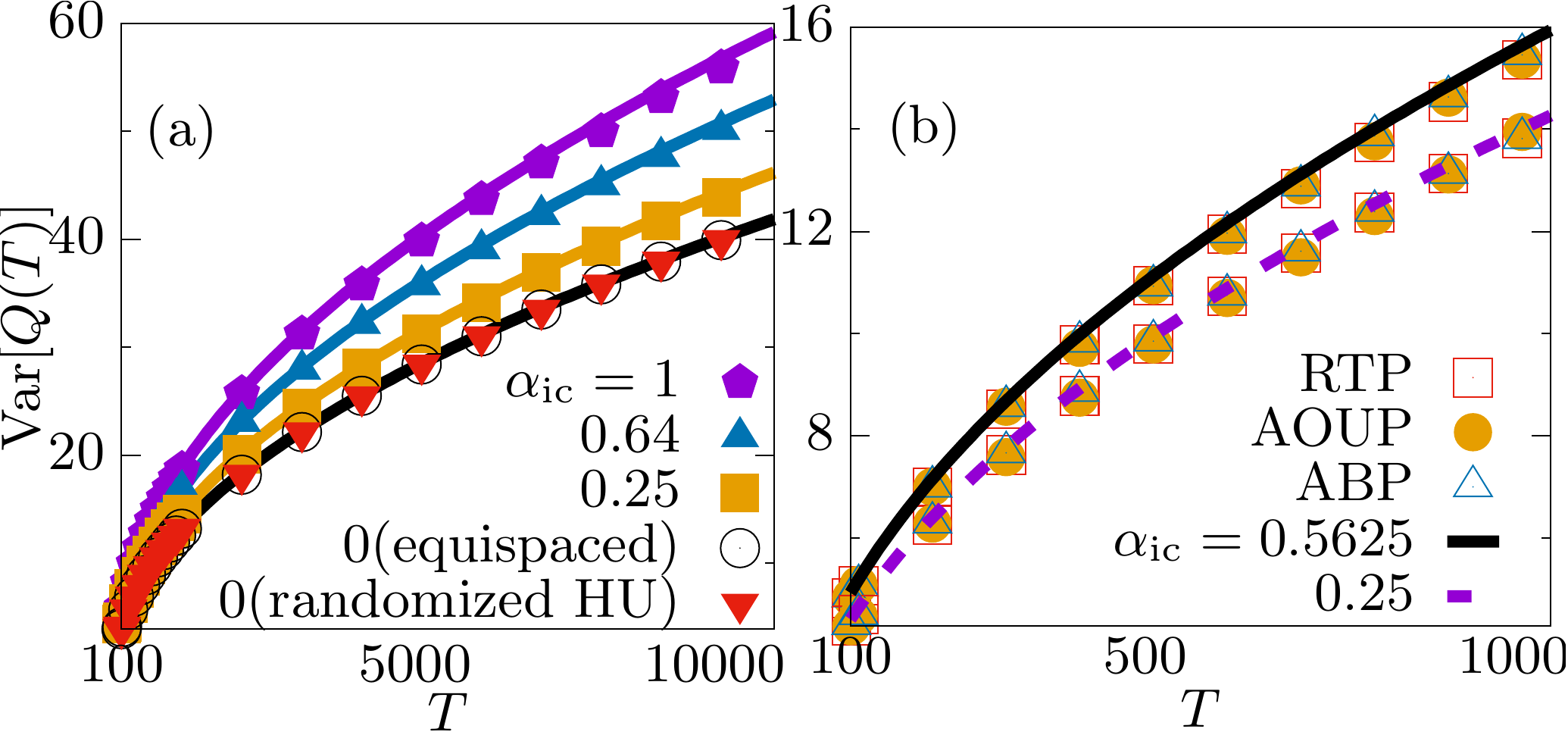}
\caption{
(a) The variance $\operatorname{Var}[Q(T)]$ of non-interacting Brownian particles with $D=1$
and $\rhobar=1$, at different values of $\alpha_{\rm ic}$.  The MC simulation results (points) match the theoretical prediction~(\ref{variance-nonint-intro}) (solid lines). 
For HU initial states ($\alpha_{\rm ic}=0$), two different initial setups are shown: equi-spaced initial positions {with or without additional random displacements}.
(b) The variance ${\rm Var}[Q(T)]$ for noninteracting active particles at long times. Points show simulation results;
lines are the prediction \eqref{variance-nonint-intro}.
}\label{Variance-plots}
\end{figure}

{Figure}~\ref{Variance-plots}(a) shows $\operatorname{Var}[Q(T)]$ for point Brownian particles in 1D, obtained from Monte Carlo (MC) simulations for various $\alpha_{\rm ic}$. {In the initial state for these numerics, particles are placed at random, with spacings constrained to exceed} some constant $r_0$: this yields $\alpha_{\rm ic}=(1-r_0\rhobar)^2$, providing a family of initialization protocols spanning $\alpha_{\rm ic}$ between $0$ ({equal} spacing $r_0=1/\rhobar$) and unity (ideal gas, $r_0=0$)~\cite{SM}.   All results match equations (\ref{variance-nonint-intro},\ref{V2-fin},\ref{V1-fin}).
To {check} that all dependence on the initial conditions comes from $\alpha_{\rm ic}$, we also simulated a different HU initial ensemble where equi-spaced particles receive independent random displacements of fixed size~\cite{SM}: the long-time behavior matches that for equi-spaced initial particles. 

{Figure}~\ref{Variance-plots}(b) shows results for three popular models of active particles~\cite{ion-urna-active-gaussian, aoups-martin, abp-review, aoups-paper,abp-paper,rtp-paper,rtp-Christ, cates-tailleur-epl}: active Ornstein-Uhlenbeck particles (AOUPs), active Brownian particles (ABPs) and run-and-tumble particles (RTPs), see~\cite{SM} for details.   These systems all satisfy (\ref{Uzt-Brownian}) {with $D$ their late-time diffusivity}.  They {precisely obey our} predictions (\ref{variance-nonint-intro},\ref{V2-fin},\ref{V1-fin}). 

Notably, Eq.~(\ref{variance-nonint-intro}) also applies in higher dimensions, to fluctuations of the flux passing through a planar boundary: it suffices that the particles' normal distances from the boundary are independent, Markovian, and obey~\eqref{Uzt-Brownian}.
Moreover, 
the current in a system of hard Brownian point particles undergoing single-file motion has the same statistics as in the non-interacting case~\cite{BJC-cambridge}. Hence (\ref{variance-nonint-intro}) also applies in that single-file system, which we address next.

\begin{figure}
\includegraphics[width=\columnwidth]{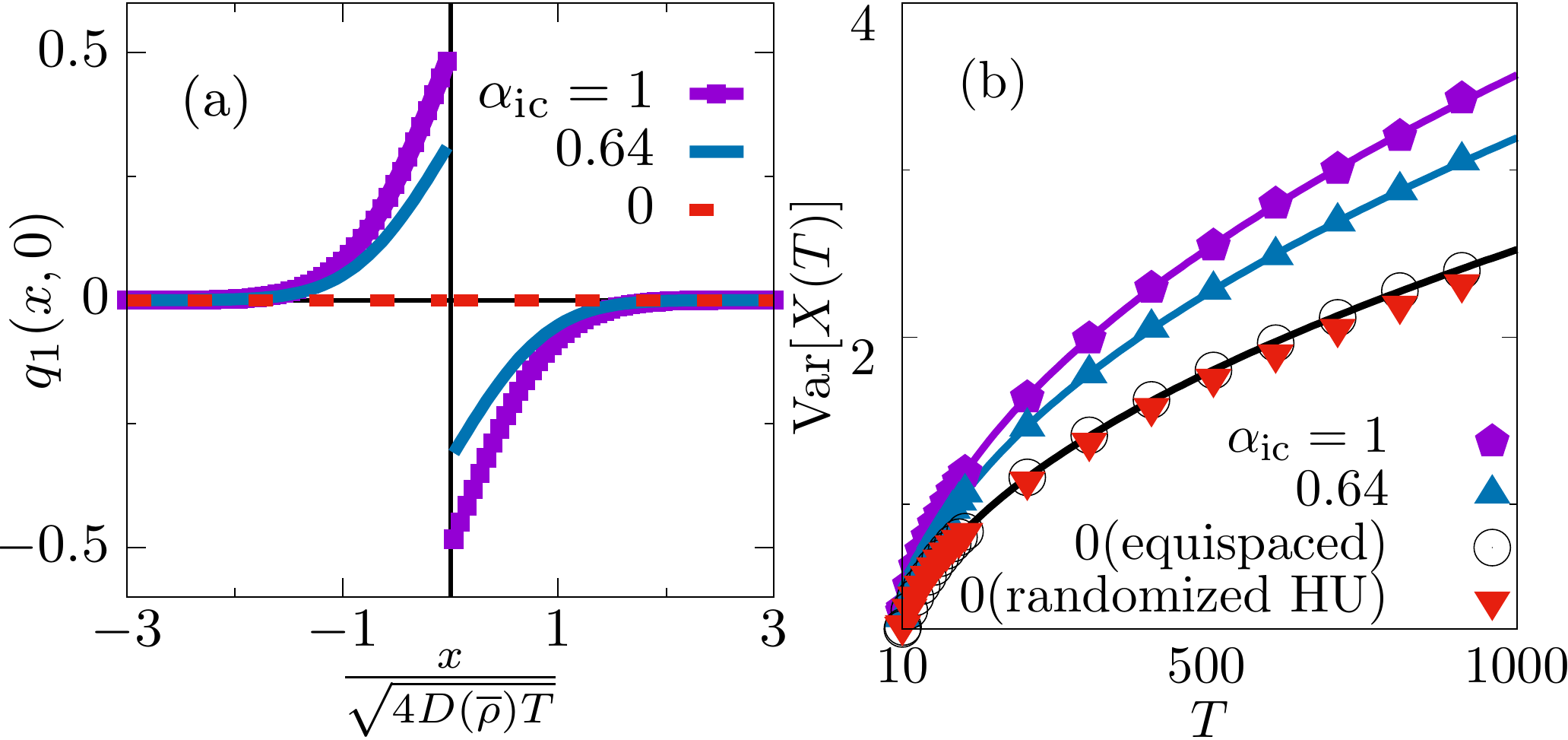}
\caption{
{(a)~The non-typical initial condition $q_1(x,0)$ that appears when considering fluctuations of {$X(T)$} for various $\alpha_{\rm ic}$.  Large currents correspond to an excess of particles to the left of the origin, {biasing} the tracer to the right.
(b)~The mean-squared tracer displacement $\operatorname{Var}[X(T)]$ in a system of point Brownian particles at density $\rhobar=10$.  
Numerical results are shown as {points; solid lines show} the theoretical prediction~(\ref{VarX-point}). 
We show results for two different HU initial conditions, {with initialisation protocols the} same as Fig.~\ref{Variance-plots}.
}
}
\label{VarX-pph}
\end{figure}

{
{\em Single-file tracer motion:}  We now consider an infinite {1D system of diffusive, hardcore particles}, whose initial condition {is homogeneous, with mean} density $\rhobar$ and two-point correlation $C(z)$.  A single tracer particle is identified, and its displacement between times $0$ and $T$ is denoted by $X(T)$, whose variance obeys (\ref{variance-tracer-intro}), as we now show.  The physical mechanism for this result is the same as that leading to (\ref{variance-nonint-intro}), although the computation is more involved. The method follows previous work~\cite{sadhu-prl}; we outline it here, with details in~\cite{SM}.

We first assume that the hydrodynamic density field $\rho$ obeys the MFT equation~\cite{MFT-main}
\begin{equation}\label{MFT-dynamics}
\partial_{t}\rho(x,t) = \partial_{x} [D(\rho) \partial_x \rho(x,t) + \sqrt{\sigma(\rho)} \eta(x,t)],
\end{equation}
where $\sigma(\rho)$ and $D(\rho)$ are the mobility and diffusivity {and $\eta(x,t)$ is Gaussian spatiotemporal white noise.  Examples of such MFT systems include hard Brownian particles~\cite{sadhu-prl}, and the symmetric simple exclusion process~\cite{mallick-mft-sep-exact}.}

The moment generating function (MGF) of the tracer position is $\langle e^{\lambda X(T)} \rangle$, which can be expressed  as a path integral in the Martin-Siggia-Rose formalism~\cite{sadhu-prl}: 
\begin{equation}\label{mgf-1}
\langle e^{\lambda X(T)} \rangle = \int \mathscr{D} [\rho(x,t), \hat{\rho}(x,t)] \, e^{-\mathcal{S}[\rho(x,t), \hat{\rho}(x,t)]},
\end{equation}
where the average $\langle ... \rangle$ now includes both the random initial condition and the stochastic dynamics of the density;  $\hat{\rho}(x,t)$ is a response field, and the action is
\begin{equation}
\mathcal{S}[\rho, \hat{\rho}] = -\lambda X(T) + F[\rho]  + \int_0^T \!dt \int_{-\infty}^\infty \!\!dx\, {\cal L}(\rho,\hat\rho)\,.
\end{equation}
{Here} $F[\rho]$ is the log-probability of the initial condition, and
\begin{equation}
 {\cal L}(\rho,\hat\rho)
= 
\hat{\rho} \partial_{t}\rho 
 - \frac{\sigma(\rho)}{2} (\partial_x \hat{\rho})^2 + D(\rho) (\partial_x \rho) (\partial_x \hat{\rho}) \,.
 \end{equation}
For single-file motion, $X(T)$ is fully determined by the dynamics of the density: at time $T$, all particles between the tracer and the origin must have had initial {positions} $y_i<0$.  This implies~\cite{sadhu-prl,SM}
\begin{equation}\label{conservation1}
\int_0^{X(T)} \!dx\,  \rho(x,T) = \int_0^{\infty} \!dx\,   [\rho(x,T) - \rho(x,0)]\, \,.
\end{equation}

On hydrodynamic time scales, {noise  becomes weak and} (\ref{mgf-1}) can be evaluated by a saddle-point method.  Physically, this involves the computation of an instanton that generates a large tracer displacement $X(T)$, whose size is determined by the parameter $\lambda$.  To find the variance of $X(T)$,  the computation is required to $O(\lambda^2)$.  At this level, the instanton dynamics is $\rho(x,t) \approx \rhobar + \lambda q_1(x,t)$ and $\hat\rho(x,t) \approx \lambda p_1(x,t)$, where $p_1,q_1$ are canonically conjugate fields; terms at higher order in $\lambda$ can be neglected.  

Within MFT, the log-probability of the initial state is determined by a function $g_{\rm ic}$, as
$F[\rho] = \int_{-\infty}^\infty \!dx\, g_{\rm ic}(\rho(x,0))$, specifying the probability of local density fluctuations~\cite{SM,MFT-main}.  
We emphasise {that} $\rho$ is the hydrodynamic density: there may be density correlations on the {scale of} the interparticle spacing, but $g_{\rm ic}$ is still a local function of $\rho$. 
Since the density fluctuations are of order $\lambda$,
it is consistent to approximate~\cite{SM,MFT-main}
\begin{equation}
g_{\rm ic}(\rho) \approx   [\rho(x,0) - \rhobar]^2  / (2 \alpha_{\rm ic} \rhobar )
\end{equation}
where $\alpha_{\rm ic}=1/(\rhobar g''_{\rm ic}(\rhobar))$ is the {Fano factor} defined in (\ref{fano}). 

Two special cases are relevant: first, if the initial condition has no fluctuations then $\rho(x,0)=\rhobar$ exactly and $\alpha_{\rm ic}\to0$.  The corresponding result for $\operatorname{Var}[X(T)]$ coincides with the `quenched' case addressed in~\cite{sadhu-prl}.  Second, if the initial condition is a thermally equilibrated steady state of (\ref{MFT-dynamics}) then a fluctuation-dissipation theorem requires
$g''_{\rm ic}(\rhobar)=2D(\rhobar)/\sigma(\rhobar)$~\cite{MFT-main}.  
In this case, $g_{\rm ic}$ follows from the model's free energy, and $\operatorname{Var}[X(T)]$ coincides with the `annealed' variance of~\cite{sadhu-prl}.    Hence (as for the noninteracting particles considered above) our formalism incorporates these two special cases, but extends them to arbitrary initial protocols with no assumption of thermal equilibration.

The instanton dynamics is obtained by extremising the action, leading to
\begin{eqnarray}
\partial_{t} q_1(x,t) & =&  \partial_x [ D(\rhobar) \partial_{x} q_1(x,t)  -\sigma(\rhobar)\, \partial_{x} p_1(x,t) ] \label{q1-evol} 
\nonumber \\
\partial_{t} p_1(x,t) & = & - D(\rhobar) \partial_{xx} p_1(x,t) \, \, , \label{p1-evol}
\end{eqnarray}
with boundary conditions~\cite{SM}
\begin{eqnarray}
q_1(x,0) &=& \rhobar\, \alpha_{\rm ic} [p_1(x,0) - p_1(x,T)] \label{q1-bc-genf} \\
p_1(x,T) &=& \Theta(x)/\rhobar \label{p1-bc-genf}\,.
\end{eqnarray}
The equations for $p_1$ are closed and {exactly solvable, following}~\cite{sadhu-prl}.  
Hence {(\ref{q1-bc-genf}) sets} the initial condition for the instanton, which is the fluctuation of the initial condition {$\rho(x,0)$ associated} with the prescribed fluctuation of $X(T)$: 
\begin{equation}
q_1(x,0) = \alpha_{\rm ic} \left[ \frac12 {\rm erfc} \left(\frac{-x}{\sqrt{4 D(\rhobar) T }} \right) - \Theta(x) \right] .
\end{equation} 

This result is shown in Fig.~\ref{VarX-pph}(a). {In physical terms}, if the initial density is large on the left of the tracer, then it tends to move to the right, and vice versa.  The size of the fluctuation is set by $\alpha_{\rm ic}$ {(vanishing for HU initial conditions, which lack these large-scale density fluctuations by definition)}; the associated length scale is $\sqrt{4D(\rhobar)T}$.  Like the {current fluctuations considered earlier, this} shows that fluctuations of tracer position are strongly coupled to the number of particles within this distance {on either side} of the origin.}

{
Finally, 
the variance of the tracer position is set by the second derivative of the MGF.  We thereby obtain (\ref{variance-tracer-intro}) with
\begin{align}
\Delta X^2_{\rm noise}(T) & {\,\simeq\,} \frac{1}{\rhobar} \sqrt{\frac{\sigma(\rhobar)^2 T}{2 \pi  \rhobar^2 D(\rhobar)}} \;  \label{deltax2},
\\ 
\Delta X^2_{\rm dens}(T) & {\,\simeq\,}  \frac{\sqrt{2}-1}{ \rhobar} \sqrt{\frac{2 D(\rhobar) T}{\pi}} \, . \label{deltax1}
\end{align}
Once again, $\alpha_{\rm ic} = 0$ corresponds to the `quenched' result given in \cite{sadhu-prl}, while their annealed result is recovered for the thermally equilibrated value of $\alpha_{\rm ic}$.

For the specific case of hardcore Brownian particles, 
one {also} has $\sigma(\rhobar)=2 D \rhobar$ and $D(\rhobar) = D$. Hence,
\begin{equation}\label{VarX-point}
\operatorname{Var}[X(T)]  {\,\simeq\,} \frac{1}{\rhobar} \sqrt{\frac{2 D T}{\pi}} \left[1+ \alpha_{\rm ic} (\sqrt{2}-1) \right].
\end{equation}
This result is verified numerically in Fig.~\ref{VarX-pph}(b), for three different values of $\alpha_{\rm ic}$ with $D=1, \rhobar=10$.  For the HU case ($\alpha_{\rm ic}=0$), we also show that two different initial preparations yield the same {result, similarly to Fig.~\ref{Variance-plots}}. 

To summarise, we have considered two different situations where the initial preparation of a {1D} diffusive system has long-lasting effects on its fluctuating dynamics.   We showed that this dependence {is} captured by the single parameter $\alpha_{\rm ic}$, which quantifies large-scale density fluctuations {in the initial state}. 
The resulting framework generalizes previous results to a vastly wider range of initialization protocols.  The coupling of $\alpha_{\rm ic}$ to the long-time dynamics occurs because $X(T)$ and $Q(T)$ are both correlated with the number of particles that were initially within a distance $\sqrt{4DT}$ of the origin: for large $T$, the fluctuations in this number are set by $\alpha_{\rm ic}$.
We suspect that similar effects arise in several other systems~\cite{kundu-cividini,rana-sadhu-arxiv, krapivsky-meerson} where initialisation protocols have  everlasting effects (see also~\cite{dandekar-mallick-arxiv}).

These results may be important for quantitative experiments on single-file diffusion, such as NMR measurements inside molecular sieves~\cite{zeolite, karger-zeolite}.  Eq.~(\ref{VarX-point}) predicts that the molecular transport ratio $\operatorname{Var}[X(T)]/\sqrt{T}$ has an everlasting dependence on $\alpha_{\rm ic}$, which itself depends on the (typically nonequilibrium) conditions under which the molecules are loaded into the sieve~\cite{zeolite, karger-zeolite}. Without controlling for this dependence, measurements of this ratio may not yield reproducible results. Our predictions could also be tested in colloidal systems, following~\cite{kollman-bechinger}. We also note that our results on single-file motion could lead to interesting consequences for related harmonization studies on bead-spring systems~\cite{Lizana-pre}.

We have emphasized the special role of HU initial states, whose long-wavelength density fluctuations have vanishing amplitude.  All such states lie within a single universality class that also contains `quenched' systems~\cite{derr-gers,sadhu-prl,tirtha-paris}.  While many previous works focussed on their creation ~\cite{tjhung-berthier-prl, hexner-levine-prl, torquato-stealth}, our work reveals a new dynamical {\em consequence} of HU states
(see also~\cite{lee-jstat-mech-2019, kwon-kim-pre}).

While we have analysed the variances of dynamical quantities, their {higher moments} (and large deviations) presumably depend on higher-order statistics of the initial state, {suggesting} rich new possibilities for future research. Further open questions arise for single-file systems not described by MFT~\cite{BJC-cambridge}; for driven 1D systems~\cite{satya-mustansir,van-beijeren, barkai-silbey,Masi-Ferrari,Rajesh-Satya,Mallmin-Blythe-Evans,illien-biased}; and in higher dimensions~\cite{liu-nanoscale,transport-anisotropy,FR-anisotropy}.

\if{
\vspace{12pt}
\noindent\rlj{\emph{RLJ: Some insights that might get dropped back in if we think they are sufficiently important}}:

\rlj{1. We emphasize that for an initial state characterised by $\alpha_{\rm ic}$, one may introduce a finite waiting time between preparation and the start of the dynamical measurement without affecting the asymptotic results (\ref{variance-tracer-intro},\ref{variance-nonint-intro}), because the value of $\alpha_{\rm ic}$ will not change during any finite waiting time.}

\rlj{2. Note that previous work~\cite{kollmann-prl,sadhu-prl,rana-sadhu-arxiv} identified relationships between compressibility and the tracer MSDs, but the derivations are restricted to equilibrated initial conditions, through the FDT.  Our results show that the fundamental relationship is between the dynamical quantities and the Fano factor $\alpha_{\rm ic}$ of the initial condition, and does not rely on any properties of equilibrium systems.}

\fi

{\em Acknowledgments:} {We thank} Tal Agranov, Camille Scalliet, Johannes Pausch and Jean-Fran\c{c}ois Derivaux for helpful discussions. Work funded in part by the European Research Council under the Horizon 2020 Programme, ERC Grant Agreement No. 740269. MEC is funded by the Royal Society. 




\end{document}


\title{Supplemental Material for ``Role of initial conditions in $1D$ diffusive systems: compressibility, hyperuniformity and long-term memory"}
\author{Tirthankar Banerjee}
\affiliation{DAMTP, Centre for Mathematical Sciences, University of Cambridge, Wilberforce Road, Cambridge, CB3 0WA}
\author{Robert L. Jack}
\affiliation{DAMTP, Centre for Mathematical Sciences, University of Cambridge, Wilberforce Road, Cambridge, CB3 0WA}
\affiliation{Yusuf Hamied Department of Chemistry, University of Cambridge, Lensfield Road, Cambridge CB2 1EW, United Kingdom}

\author{Michael E. Cates}
\affiliation{DAMTP, Centre for Mathematical Sciences, University of Cambridge, Wilberforce Road, Cambridge, CB3 0WA}

\maketitle

\newcommand{\rlj}[1]{{\color{blue}#1}}

In the following we present some supplemental information, to provide additional detail on the discussion of the main text.  Sec.~\ref{sec:qu-ann} gives some theoretical background on the meaning of quenched and annealed variances.  In Sec.~\ref{compress_analytical} we describe a set of translationally invariant initial states with Fano factor $\alpha_{\rm ic} \leq 1$, which are used as initial conditions for the numerical computations in the main text.  Section \ref{numerics-active} contains additional details of the numerical results. In  Sec.~\ref{mft-analysis}, we give a full derivation of the tracer variance for systems described by Macroscopic Fluctuation Theory (MFT).  This computation is included for completeness, it largely follows Ref \cite{sadhu-prl}, with some modifications to account for the general class of initial conditions considered in this work.

{
\section{Quenched and annealed averages}
\label{sec:qu-ann}

This section gives full definitions of the averaging procedures that are used in the main text, and we explain the terms ``quenched variance'' and ``annealed variance''.  We focus on the case of current fluctuations for non-interacting particles, the extension to tracer diffusion in the MFT setting is immediate.  

Consider a system with a random initial condition $\mathbf{y}$ whose probability distribution is $P_{\rm ic}(\mathbf{y})$.  The system's dynamics are also stochastic, and its (random) configuration at time $t$ is $\mathbf{x}_t$.  It is natural to define the conditional distribution $P_{\rm dyn}(\mathbf{x}_t|\mathbf{y})$ which is the probability distribution for $\mathbf{x}_t$, given that the system started in a specific state $\mathbf{y}$.  In our setting, $\mathbf{y}$ indicates the initial particle positions, and $\mathbf{x}_t$ their positions at time $t$.

Now write $Q(t)=\mathcal{Q}(\mathbf{x}_t)$ for some observable quantity in the system at time $t$ (which in our case is the current that has passed through the origin).  For a given initial condition $\mathbf{y}$, the average of $Q(t)$ is
\beq
\langle Q(t) \rangle_{\mathbf y}  = \int \mathcal{Q}(\mathbf{x}_t) P_{\rm dyn}(\mathbf{x}_t|\mathbf{y}) d\mathbf{x}_t .
\eeq
This is the definition of the $\langle \ldots \rangle_{\mathbf y}$ averaging procedure.

Similarly, for any function of the initial state, $A=A(\mathbf{y})$ the definition of the $\overline{(...)}$ average is
\beq
\overline{A} = \int A(\mathbf{y}) P_{\rm ic}(\mathbf{y}) \, d \mathbf{y}
\eeq
Moreover, since $\langle Q(t) \rangle_{\mathbf y}$ is itself a function of the initial condition, the full average (which accounts for the randomness of both the initial condition and the dynamics) is
\beq
\overline{ \langle Q(t) \rangle_{\mathbf y} }  = \int \int \mathcal{Q}(\mathbf{x}_t) P_{\rm dyn}(\mathbf{x}_t|\mathbf{y})  P_{\rm ic}(\mathbf{y}) d\mathbf{x}_t d\mathbf{y} .
\eeq
Such full averages are denoted as
\beq
\langle Q(t) \rangle = \overline{ \langle Q(t) \rangle_{\mathbf y} } 
\eeq
For the mean, this double average is simple, but the situation is more complicated when one considers variances, or higher moments.  

Here we consider the variance, which is sufficient for our purposes.  By analogy with the full average defined above, we consider the full variance, which again accounts for the randomness of both the initial condition and the dynamics.  This is
\beq
\operatorname{Var}[Q(t)] = \big\langle Q(t)^2 \big\rangle - \big\langle Q(t) \big\rangle^2
\eeq
As a thought experiment to measure the full variance, one should perform many experiments, each with a different (random) initial condition, and measure the variance of $Q(t)$ among the outcomes.

On the other hand, one may also consider the fluctuations of $Q_t$ that occur if one fixes the initial condition, but includes the randomness of the dynamics.  This is a $\mathbf{y}$-dependent variance, that is
\beq
\delta Q(t,\mathbf{y})^2 = \big\langle Q(t)^2 \big\rangle_\mathbf{y} - \big\langle Q(t) \big\rangle_\mathbf{y}^2
\eeq
The thought experiment for measuring $\delta Q(t,\mathbf{y})^2$ is to perform many experiments, but always with the same initial condition $\mathbf{y}$.  Then measure the variance of $Q$ among these outcomes.

To avoid dealing with specific initial conditions, one may now average this quantity over the initial condition, to obtain what is known as the quenched variance:
\begin{align}
\Delta Q_{\rm noise}(t)^2 & = \overline{ \delta Q(t,\mathbf{y})^2 } 
\nonumber\\ &
= \overline{\big\langle Q(t)^2 \big\rangle_\mathbf{y}} -  \overline{\big\langle Q(t) \big\rangle_\mathbf{y}^2 } 
\end{align}
In most situations -- including those discussed here -- the quenched variance can be interpreted as the value of $\delta Q(t,\mathbf{y})^2 $ that would be obtained for a typical initial condition $\mathbf{y}$.   The corresponding annealed variance is $\operatorname{Var}[Q(t)]$.

Now observe that 
\begin{align}
\operatorname{Var}[Q(t)] & = \overline{\big\langle Q(t)^2 \big\rangle_\mathbf{y}} - \overline{\big\langle Q(t) \big\rangle_\mathbf{y}}^2
\nonumber \\
& =  \overline{\big\langle Q(t)^2 \big\rangle_\mathbf{y}} - \overline{\big\langle Q(t) \big\rangle_\mathbf{y}^2 }+\overline{\big\langle Q(t) \big\rangle_\mathbf{y}^2 } - \overline{\big\langle Q(t) \big\rangle_\mathbf{y}}^2
\nonumber \\
& = \Delta Q_{\rm noise}(t)^2  + \Delta Q_{\rm ic}(t)^2
\label{equ:QQQ}
\end{align}
where the first line is the definition of the full average, the second simply adds and subtracts $\overline{\langle Q(t) \rangle_\mathbf{y}^2 }$ on the right hand side, and we introduced
\beq
 \Delta Q_{\rm ic}(t)^2 =  \overline{\big\langle Q(t) \big\rangle_\mathbf{y}^2 } - \overline{\big\langle Q(t) \big\rangle_\mathbf{y}}^2 
 \label{equ:Qic}
\eeq
Hence we have recovered (8) of the main text.

One may also identify $ \Delta Q_{\rm ic}(t)^2$ as a variance by rearranging terms as 
\beq
\Delta Q_{\rm ic}(t)^2 =  \overline{ \left( \big\langle Q(t) \big\rangle_\mathbf{y} - \overline{\big\langle Q(t) \big\rangle_\mathbf{y}}\right)^2 } .
\eeq
Since this quantity is non-negative, \eqref{equ:QQQ} shows that $\operatorname{Var}[Q(t)] \geq \Delta Q_{\rm noise}(t)^2$: the annealed variance is at least as big as the quenched variance.

For ergodic physical systems without long-term memory, the behaviour at long times should become independent of their initial state.  That is, $\langle Q(t) \rangle_{\mathbf{y}} \simeq \langle Q(t) \rangle$ at long times, independent of $\mathbf{y}$.  In that case one sees from \eqref{equ:Qic} that  $\Delta Q_{\rm ic}(t)^2\simeq 0$ and the quenched and annealed averages coincide, that is $\operatorname{Var}[Q(t)]  \simeq \Delta Q_{\rm noise}(t)^2 $.   This is the sense in which everlasting differences between quenched and annealed variances point to the existence of long-term memory.

As a final remark on context (and to make direct contact with \cite{derr-gers}), note that the ``quenched/annealed'' nomenclature comes from the context of disordered systems.  The role of disorder in that case is played here by the initial condition.  For disordered systems, the quenched variance measures the size of fluctuations between different measurements performed on a single (typical) sample.  The annealed variance measures the size of fluctuations if a different sample is used for each measurement.  In disordered systems, this distinction is more commonly drawn at the level of the free energy, which in our context would be a cumulant generating function.  Specifically, it can be verified from the definitions of the various averages that the quenched variance of $Q$ obeys
\beq
\Delta Q_{\rm noise}(t)^2 = \left. \frac{\partial^2}{\partial \lambda^2} \overline{ \log \left\langle {\rm e}^{\lambda Q(t)} \right\rangle_\mathbf{y} }\right|_{\lambda=0}
\eeq
while the full variance obeys
\beq
\operatorname{Var}[Q(t)] = \left. \frac{\partial^2}{\partial \lambda^2}  \log \overline{\left\langle {\rm e}^{\lambda Q(t)} \right\rangle_\mathbf{y} }\right|_{\lambda=0} \; .
\eeq
}

\section{Microscopic example for the Fano factor $\alpha_{\rm ic}$: hard rods in one dimension}\label{compress_analytical}

\subsection{Theory}

A central role in this work is played by the Fano factor $\alpha_{\rm ic}$.  As an example of an equilibrium system where this quantity can be varied continuously between $0$ and $1$, we consider hard particles of size $r_0$, in one dimension.  For any configuration of $N$ such particles in a box of size $L$, there is a corresponding configuration of point particles in a box of size $L-Nr_0$.  The two configurations are related by ordering the positions of the point particles as $\tilde{x}_1,\tilde{x}_2,\dots,\tilde{x}_N$, and then defining the positions of the hard particles as $x_m = \tilde{x}_m + (m-1)r_0$.  Clearly $r_0\leq L/N$ in order that the hard particles can fit inside the box.

Since this transformation has unit Jacobian, the configurational partition function for the hard rods is the same as that of the point particles, that is
\begin{equation}\label{Part-func}
\mathcal{Z}
= \frac{(L-Nr_0)^N}{N!} \, \, .
\end{equation}
and the corresponding pressure $P$ is given by
\begin{equation}\label{pressure}
\beta P= \frac{\partial }{\partial L} (\ln Z) =  \frac{N }{L-Nr_0}  \, \, ,
\end{equation}
where $\beta$ is the inverse temperature (divided by Boltzmann's constant).

Recall that the (mean) density is $\rhobar=N/L$. Observe that $0\leq \rhobar r_0 \leq 1$, where the lower limit corresponds to point particles (hard rods of size zero) and the upper limit to the case where the hard rods fill the box exactly.  It is convenient to define a dimensionless quantity:
\begin{equation}\label{zeta}
\zeta 
= \frac{L-Nr_0}{Nr_0} 
= \frac{1 - \rhobar r_0}{\rhobar r_0}   \, \, ,
\end{equation}
which diverges for point particles, and approaches zero for the case where the rods fill the box.

 From \eqref{pressure}, we can calculate the (isothermal) compressibility $\kappa$ as
\begin{equation}\label{alp-2}
\kappa = -\frac{1}{L}\frac{\partial L}{\partial P} 
= 
\frac{\beta (1-\rhobar r_0)^2}{\rhobar} \, \, .
\end{equation}
Then it is well-known (by considering large systems in the grand canonical ensemble~\cite{chaikin-lubensky, bell-fcthm}) that the Fano factor for fluctuations of the particle number is 
%
\begin{equation}
\alpha_{\rm ic} =  \frac{ {\rm Var}(n) }{ \langle n \rangle} = \frac{\kappa \rhobar }{\beta}
\end{equation}
(The result for $\alpha_{\rm ic}$ is the same if one considers a large grand-canonical system, or a large subsystem of size $\ell$ in a very large canonical system.)
For the present system \eqref{zeta} and \eqref{alp-2} yield
\begin{equation}\label{alpD-result}
\alpha_{\rm ic} = 
 \left( \frac{\zeta}{1+\zeta} \right)^2 \, \, .
\end{equation}
Since $0\leq\zeta<\infty$, one sees that these states have $0\leq \alpha_{\rm ic} \leq 1$.

\begin{figure}
\includegraphics[width=8cm]{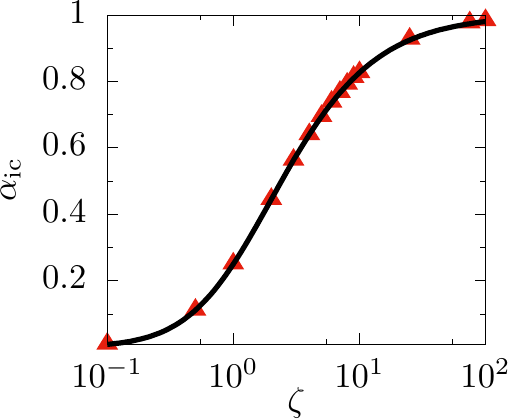}
\caption{Numerically estimated $\alpha_{\rm ic}$ from hard rod configurations with varying $\zeta$.  The line shows the prediction (\ref{alpD-result}).  Results were obtained at density $\rhobar=1$ in a system of $N=10000$ particles.
}\label{alphaD-an-num-comp}
\end{figure}

{

\subsection{Numerical generation of states with prescribed $\alpha_{\rm ic}$}

The above analysis provides a method for generating states with prescribed $\alpha_{\rm ic}$.  This method was used in the simulations described in the main text.   Specifically: for any $\rhobar,\alpha_{\rm ic}$, Eqs.~(\ref{zeta},\ref{alpD-result}) can be solved for the corresponding $r_0$.  
Then, for initial states with all particles to the left of the origin (as required for numerical computations of the $Q(T)$), we start by placing $N$ point particles at random between $-L$ and $-Nr_0$.  Similar to the construction above, these positions can be ordered in an increasing sequence as $\tilde{x}_1,\dots,\tilde{x}_N$ and the position of the $m$th hard particle is obtained as $x_m = \tilde{x}_m + (m-1)r_0$.  

For initial states with particles on both sides of the origin (as required for simulations of tracer motion), we start with the same method identified above.  Then we identify the central particle as the tracer, and we shift the whole configuration so that the tracer starts at the origin.

To confirm that these methods work as expected, we generated configurations in this way.  Measuring the asymptotic variance $\alpha_{\rm ic}$ numerically requires some care because the number of particles is finite in the simulations, which are performed in the canonical ensemble.
Let $n(\ell)$ denote the number of particles within a distance $\ell$ of the origin.  We measured the mean and variance of this number, as a function of $\ell$ .  By homogeneity, $\langle n(\ell) \rangle = \rhobar \ell$.  Also, the variance is found to be $\operatorname{Var}[n(\ell)] \simeq \alpha \ell(L-\ell) N /L^2$ [valid for large $N,L$ with $N/L=\rhobar$ and $\ell\gg 1$].  

Note in particular $\operatorname{Var}[n(L)]=0$ because the total number of particles is fixed.  To obtain $\alpha_{\rm ic}$, we note that its definition includes a limit
of large $\ell$, which must be taken \emph{after} the thermodynamic limit $L\to\infty$ (at fixed $\rhobar$).  That is
$$
\alpha_{\rm ic} = \lim_{\ell\to\infty} \lim_{L\to\infty} \frac{ {\rm Var}[n(\ell)] }{ \langle n(\ell) \rangle} = \alpha
$$  
Since $\ell\ll L$, this result is independent of the ensemble.  
In practice we take configurations from the canonical ensembles as above (fixed $N$) and we fit numerical data for $\operatorname{Var}[n(\ell)]$ to the quadratic form $\alpha \rhobar \ell(L-\ell)/L $ in order to obtain $\alpha$ as an estimate of $\alpha_{\rm ic}$.  We show in Fig.~\ref{alphaD-an-num-comp} that the results agree perfectly with the prediction (\ref{alpD-result}).

Note that states with $\alpha_{\rm ic} > 1$ can be generated in principle, by first distributing all point particles on the line uniformly for a given $\rhobar$, adding suitable local attractive interactions between particles, and equilibrating the system. We do not consider such states in our numerical computations.  


}

\section{Numerical schemes : further details}\label{numerics-active}

We  briefly describe additional numerical schemes used to obtain the numerical results in the main text (Figs.\ 1 and 2) .

\subsection{Hyperuniform initial states}

{For hard-rod configurations, there is a hyperuniform (HU) state where $L=Nr_0$, so all particles are touching their neighbours.  The particles are equispaced, the system resembles a crystal, and $\alpha_{\rm ic}=0$.  
As described in the main text, results that depend only on $\alpha_{\rm ic}$ must give the same result for any HU initial condition.

To test this, we modify these equispaced configurations by short-ranged perturbations, which do not disturb the long wavelength density fluctuations, and hence preserve the HU property.  Specifically, we add independent and identically distributed random increments to each particle position: the increment is $\pm\delta$ where positive and negative signs are each chosen with probability $1/2$.  (For the case where all particles are initialised to the left of the origin, particles within $\delta$ of the origin are we always incremented by $-\delta$.) For the ``randomized HU'' initial conditions with $\alpha_{\rm ic}=0$ shown in Fig.~1(a) of the main text, the density is $\rhobar=1$ and we take $\delta=5$, so particles are displaced by several times their typical spacing.  For Fig.~2(b) we have similarly $\rhobar=10$ and $\delta=1$.}


\subsection{Dynamics}
For the non-interacting problem we study four different models, all of which have late-time Gaussian statistics. We list these in the following:
\vspace{0.3cm}


{\em Passive Brownian motion:} Particle $i$ undergoes Brownian motion according to
\begin{equation}\label{pass-Brow}
\dot{x}_i = \sqrt{2 D} \eta_{i}(t),
\end{equation}
where $D$ is the diffusivity, and $\eta_i(t)$ is a zero-mean white noise with a variance, $\langle \eta_i(t) \eta_j(t') \rangle =\delta(t-t') \delta_{ij}$. The variance of the current associated with this system of particles, starting from a step-like initial set-up and different values of $\alpha_{\rm ic}$ is plotted in Fig.1(a) of the main text. We took $\rhobar=1$ and $D=1$, with $N=4000$. The data was averaged over $10^6$ realizations.


\vspace{0.3cm}

{\em Run and tumble particles:} The dynamics of each run-and-tumble particle (RTP) $i$ is given by~\cite{malakar-jstat, rtp-Christ}:
\begin{equation}
\dot{x}_i = v \sigma_i(t),
\end{equation}
where 
$v$ is the velocity and $\sigma_i(t)$ is a noise that flips at a Poisson rate $\gamma$ between two values $\pm 1$. It is well known~\cite{malakar-jstat} that the late-time behaviour ($t \gg \gamma^{-1}$) of such particles is Gaussian, with an effective diffusivity $D=\frac{v^2}{2\gamma}$. In Fig.1(b), the temporal evolution of $\operatorname{Var}[Q(T)]$ for RTPs is shown by squares. We choose $v=1, \gamma=0.5, \rhobar=1, N=1000$. The data was averaged over $\sim 10^5$ samples.
\vspace{0.3cm}

{\em Active Ornstein-Uhlenbeck particles:} The equations of motion for an active Ornstein-Uhlenbeck particle (AOUP), without any external or interaction potential, are given by~\cite{aoups-martin} : 
\begin{eqnarray}\label{aoup-dyn}
\dot{x}_i &=& v_i + \sqrt{2 D_x}\eta_{i}^{x}(t)\nonumber \\
\tau\dot{v}_{i} &=& -v_i + \sqrt{2 D_v}\eta_{i}^{v}(t),
\end{eqnarray}
where $x_i$ and $v_i$, respectively are the position and propulsive velocity of particle $i$, and the $\eta_i^{a}(t)$  are zero-mean Gaussian white noises with variances $\langle \eta_i^{a}(t) \eta_j^{b}(t^{\prime}) \rangle = \delta(t-t^{\prime}) \delta_{ij} \delta_{ab}$, with $a,b=x,v$. Also, $D_x$ and $D_v$ are the noise strengths associated with $\eta_i^x(t)$ and $\eta_i^v(t)$, respectively. For $\tau=0$, we have simple passive Brownian motion, with effective diffusivity $D=D_v+D_x$. A finite $\tau$ acts as a characterisitic persistence time, thus (only) lending inertia to the particle motion. For $t \gg \tau$, each AOUP behaves diffusively, following $D=D_v+D_x$. In Fig.1(b) of the main text, we use $v_0=0, D_x=0.5=D_v, \tau=1, \rhobar=1$ and $N=1000$. Results of $\operatorname{Var}[Q(T)]$ for AOUPs (averaged over $\sim 10^5$ samples) are shown by circles in Fig.1(b).
\vspace{0.3cm}

{\em Active Brownian particles:} The Langevin equations of motion for a free active Brownian particle (ABP) $i$ in one dimension can be written as~\cite{abp-paper, abp-review} 
\begin{eqnarray}
\dot{x}_i &=& v \cos \theta_i \nonumber \\
\dot{\theta}_i &=& 2 D_R \eta_i(t),
\end{eqnarray}
where $x_i$ gives the instantaneous position of the particle, $\theta_i$ represents the particle orientation that itself follows diffusive dynamics with a rotational diffusion constant $D_R$. $\eta_i(t)$ is a delta-correlated, zero mean Gaussian white noise. Free AOUPs behave diffusively with effective diffusivity $D=\frac{v^2}{2 D_R}$ for times $t \gg D_R^{-1}$~\cite{abp-paper}. In Fig1(b) of the main text, we use $v=1, D_R=0.5$ for each ABP. The initial orientation for each particle is chosen randomly between $0$ and $2\pi$. We choose $\rhobar=1, N=1000$. The results for $\operatorname{Var}[Q(T)]$  (averaged over $\sim 10^5$ samples) are represented by triangles in Fig.1(b).
\vspace{0.3cm}

{\em Interacting hardcore Brownian particles:} Here we use $N=10000+1$ particles located within $-L/2$ and $L/2$, with $\rhobar=10$ and the tracer located centrally at the origin. We follow the technique of \cite{hegde-prl}. We let the particles evolve independently up to time $T$, and note the position of the new central particle, which equivalently gives the displacement of the tracer in a system of hard Brownian particles. The data was averaged over $\sim 10^5$ realizations. 

\section{Macroscopic fluctuation theory for the initial state with arbitrary $\alpha_{\rm ic}$}\label{mft-analysis}
We present a systematic derivation to show that the Fano factor $\alpha_{\rm ic}$ of the initial state determines the variance of tracer position $X(T)$ at some late-time $T$, for a general class of interacting particles whose dynamics is described by the framework of Macroscopic Fluctuation Theory (MFT). We follow exactly the procedure introduced in \cite{sadhu-prl}, where the analysis is set up first as a large deviation problem through the optimal solutions of the action, and then the variance $\operatorname{Var}[X(T)]$ of the tracer position is extracted by a perturbative expansion in orders of $\lambda$, which is the Laplace variable corresponding to $X(T)$ in its cumulant generating function. {It is essential that particles undergo single-file motion, in particular, the tracer cannot pass its neighbours.  We also assume that the tracer starts from the origin, so $X(T)$ can be interpreted either as its position, or its displacement.}

Within the MFT framework, the evolution of the macroscopic density $\rho(x,t)$ is given by~\cite{MFT-main,derrida-mft}
\begin{equation}
\partial_{t}\rho(x,t) = \partial_{x} [D(\rho) \partial_x \rho(x,t) + \sqrt{\sigma(\rho)} \eta(x,t)],
\end{equation}
where $\sigma(\rho)$ and $D(\rho)$ are the mobility and diffusivity, respectively, 
while $\eta(x,t)$ is a Gaussian white noise with zero mean and a variance $\langle \eta(x,t)\eta(x',t') \rangle = \delta(x-x')\delta(t-t')$.

Let $X({T})$ be the position of the tracer particle at the final (observation) time $T$. 
{Due to the single-file constraint, this position can be expressed as a functional of the density $\rho(x,t)$. 
To see this, note that the numbers of particles to left and right of the tracer must both remain constant.  At time $t$, the number of particles to the right of the tracer is $\int_{X(t)}^\infty \rho(x,t) dt$: this must be equal at times $t=0,T$, which yields}
\begin{equation}\label{conservation1-SM}
\int_0^{X(T)}  \!dx\, \rho(x,T) = \int_0^{\infty} \!dx \, [\rho(x,T) - \rho(x,0)]\,   \,.
\end{equation}
{(we used that $X_0=0$).  This formula allows the tracer position to be expressed as a functional of the density field $X(T) = X[\rho(x,T)]$. }

The statistics of $X(T)$ can be extracted from the moment generating function $\langle e^{\lambda X(T)} \rangle$, which
can be expressed as a path integral in the Martin-Siggia-Rose formalism~\cite{sadhu-prl}
\begin{equation}
\langle e^{\lambda X(T)} \rangle = \int \mathscr{D} [\rho(x,t), \hat{\rho}(x,t)] \, e^{-\mathcal{S}[\rho(x,t), \hat{\rho}(x,t)]},
\label{mgf-action}
\end{equation}
where $\hat{\rho}(x,t)$ is a response field, and the action $\mathcal{S}[\rho(x,t), \hat{\rho}(x,t)]$ is given by
\begin{equation}\label{action-tracer-SM}
\mathcal{S}[\rho(x,t), \hat{\rho}(x,t)] = -\lambda X(T) [\rho] + F[\rho(x,0)] + \int_{0}^{T} dt \int _{-\infty}^{\infty} dx \, \{ \hat{\rho} \partial_{t}\rho - \frac{\sigma(\rho)}{2} (\partial_x \hat{\rho})^2 + D(\rho) (\partial_x \rho) (\partial_x \hat{\rho}) \}.
\end{equation}
{where} $F[\rho(x,0)]=-{\rm ln} ({\rm Prob}[\rho(x,0)])$ captures the information about the initial state, with ${\rm Prob}[\rho(x,0)]$ being the probability of observing an initial density profile $\rho(x,0)$.

{

\subsection{Initial conditions and the functional $F$}

So far the analysis follows~\cite{sadhu-prl}.  However, the present derivation requires a more detailed treatment of the functional $F$, as we now discuss.
We first discuss this functional in the case where the initial condition is an equilibrium state with free energy density $f_{\rm ic}(\rho)$: we explain below how the analysis is generalised to cover non-equilibrium initial states.
{Note: this $f_{\rm ic}$ differs in general from the free energy $f_{\rm eq}$ associated with the equilibrium state of the model dynamics, which obeys the fluctuation-dissipation theorem $f_{\rm eq}''(\rho)=2 D(\rho)/\sigma(\rho)$~\cite{MFT-main}.  Taking $f_{\rm ic}=f_{\rm eq}$ will recover the annealed result of~\cite{sadhu-prl}, but the following analysis is more general, in that it applies for a general $f_{\rm ic}$, corresponding to initial conditions that are not sampled from the equilibrium stationary state of the underlying model.}

For an initial condition taken from the equilibrium state with free energy $f_{\rm ic}$, the hydrodynamic density fluctuations behave (on large length scales) as
\begin{equation}
-\ln {\rm Prob}[\rho] = F[\rho] \simeq \int_{-\infty}^{\infty} dx \, g_{\rm ic}(\rho)
\label{equ:rho-ic-ldp}
\end{equation}
with
\begin{equation}
g_{\rm ic}(\rho) = f_{\rm ic}(\rho) - f_{\rm ic}(\rhobar) - ( \rho - \rhobar )  f'_{\rm ic}(\rhobar) 
\end{equation}
as given in  Eq.~(5.1) of~\cite{MFT-main}.  
The most probable initial condition is $\rho(x,0) = \rhobar$, which has $F=0$.  
Write
\begin{align}
f_{\rm ic}(\rho) - f_{\rm ic}(\rhobar) - ( \rho - \rhobar )  f'_{\rm ic}(\rhobar)  
& =
\int_{\rhobar}^{\rho} dr [ f_{\rm ic}'(r) -  f'_{\rm ic}(\rhobar) ] 
\nonumber
\\ 
& = \int_{\rhobar}^{\rho} dr f_{\rm ic}''(\rho) [ \rho - r ] 
\end{align}
where the second equality uses an integration by parts. Hence
}
\begin{equation}\label{F_general-SM}
F[\rho(x,0)] = \int_{-\infty}^{\infty} dx \int_{\rhobar}^{\rho(x,0)} dr \, g_{\rm ic}''(r) [\rho(x,0)-r] ,
\end{equation}
{where we used $f''_{\rm ic}(\rho) = g''_{\rm ic}(\rho)$. To capture the typical fluctuations of $\rho$, it is sufficient to expand $F$ to quadratic order, yielding
\begin{equation}\label{F-quad}
F[\rho(x,0)] \approx \frac12 \int_{-\infty}^{\infty} dx \, g_{\rm ic}''(\rhobar) \, [\rho(x,0)-\rhobar]^2  \,
\end{equation}
At this order, recall that $F$ is the log-probability associated with density fluctuations in the initial state, and note that $n(\ell)=\int_0^\ell \rho(x,0) dx$ is Gaussian with mean $\ell\rhobar$ and variance
$\ell / g''_{\rm ic}(\rhobar)$, hence the Fano factor for the initial condition is
\begin{equation}\label{alp-frho-SM}
\alpha_{\rm ic} = \frac{1}{\rhobar g''_{\rm ic}(\rhobar)} \, \,.
\end{equation} 

To understand the role of initial states that are not equilibrium states, we note that the only assumption required in the following is that fluctuations have log-probability \eqref{F-quad}, where $g''_{\rm ic}(\rhobar)$ is some positive constant.  This result is assumed to hold on large (hydrodynamic) scales, so it is natural that the integrand is local.  In other words, we require that the {hydrodynamic} density has Gaussian statistics, without any long-ranged interactions. In this case the asymptotic variance $\alpha_{\rm ic}$ is necessarily related to $f''_{\rm ic}$ by  \eqref{alp-frho-SM} and we arrive at
\begin{equation}\label{F-quad-alpha}
F[\rho(x,0)] \approx \frac12 \int_{-\infty}^{\infty} \frac{dx}{\rhobar \alpha_{\rm ic} } [\rho(x,0)-\rhobar]^2  \, \, ,
\end{equation}
which will be assumed to hold at quadratic order in $\rho-\rhobar$: this is consistent with {the form of $g_{\rm ic}$ in} (\eqFF) of the main text.

}

{
\subsection{Path of least action}

As anticipated above, the strategy for computing $\operatorname{Var}[X(T)]$ is to consider large deviations of $X(T)$ as $T\to\infty$, but working to quadratic order in the size of these deviations.   We follow~\cite{sadhu-prl}, with suitable modifications to allow for a flexible choice of initial condition.  
}
{We write the cumulant generating function for $X(T)$ as
$
\mu(\lambda) = \ln \langle {\rm e}^{\lambda X(T)} \rangle .
$
This can be estimated from \eqref{mgf-action} using a saddle point method:
\begin{equation}
\mu(\lambda) = - \mathcal{S}[q(x,t), p(x,t)]
\label{equ:cgf-action}
\end{equation}
where $[\rho,\hat\rho] = [q,p]$ is the path that minimises the action $\mathcal S$.
Note that $\operatorname{Var}[X(T)]=\mu''(0)$ so it will be sufficient in the following to work to quadratic order in $\lambda$.
To find the path of least action, expand $\mathcal S$ about $(q,p)$ and set the first variation to zero: one obtains Euler-Lagrange equations}
%
\begin{eqnarray}\label{optimal-eqns-SM}
\partial_{t} q - \partial_{x} (D(q) \partial_x q) &=& -\partial_x (\sigma(q) \partial_x p) \\ \nonumber
\partial_{t} p + D(q) \partial_{xx} p &=& - \frac{1}{2} \sigma^{\prime}(q) (\partial_x p)^2 \, \,.
\end{eqnarray}

%


{The action is minimised at fixed $\lambda$ so the final position of the tracer $X(T)$ is also a variable to be optimised: denote the optimal value of this variable by {$Y(T)$}.  
The boundary condition for $t=T$ is obtained by extremising the action with respect to $\rho(x,T)$} and  using \eqref{conservation1-SM}~\cite{sadhu-prl}: 
 \begin{equation}\label{finp2-SM}
 p(x,T) = \frac{\lambda}{q(Y,T)} \Theta(x-Y) \, \,.
 \end{equation}
 where $\Theta(x)$ is the Heaviside function. 
The boundary condition at time $t=0$ is obtained similarly {by extremising the action with respect to $\rho(x,0)$}
%
\begin{equation}\label{inip-gen-SM}
p(x,0) = \frac{\lambda}{q(Y,T)} \Theta(x) + {\frac{\delta}{\delta q(x,0)}F[q(x,0)] } \, \, . %
\end{equation}

Using results obtained so far, \eqref{equ:cgf-action} becomes
\begin{equation}\label{cumulant-gen-func-SM}
\mu(\lambda) = \lambda Y - F[q(x,0)] - \int_0^{T} dt \int_{-\infty}^{\infty} dx \frac{\sigma(q(x,T))}{2} (\partial_x p(x,t))^2 \, \,.
\end{equation}

{As noted above, we are only interested in the variance of $X(T)$, so it is sufficient to make a perturbative expansion in $\lambda$ for $q(x,t)$, 
$p(x,t)$, and {$Y(T)$}.  
We work at quadratic order in $\lambda$.
For $\lambda=0$ the path of least action corresponds to the typical behaviour of the system, which is a homogeneous state $q(x,t)=\rhobar$ with typical noise realisations $p(x,t)=0$, and no net tracer motion in either direction {$Y(T)=0$}.  To leading order in $\lambda$, we have therefore}
\begin{eqnarray}
q(x,t) &=& \rhobar + \lambda q_1(x,t) 
+ \dots \label{pert1} \\
p(x,t) &=& \lambda p_1(x,t) 
+ \dots \label{pert2} \\
Y(T) & = & \lambda Y_1 + \dots
\end{eqnarray}
{Expanding \eqref{cumulant-gen-func-SM} in $\lambda$, the first terms appear at quadratic order, yielding
\begin{equation}\label{varX-F2-SM}
\frac{1}{2} \operatorname{Var}[X(T)] = Y_1 - \frac12 \int_{-\infty}^{\infty} dx \, \frac{q_1(x,0)^2 }{\rhobar \alpha_{\rm ic} } -\frac{\sigma(\rhobar)}{2} \int_0^{T} dt \int_{-\infty}^{\infty} dx \, (\partial_x p_1(x,t))^2 \, \, .
\end{equation}
where we used (\ref{F-quad-alpha}).
}

{At leading order in $\lambda$,  \eqref{conservation1-SM} becomes
\begin{equation}
\int_0^{\lambda Y_1} \rhobar dx = \int_0^{\infty} \!dx\, [ \lambda q_1(x,T) -\lambda q_1(x,0)],
\end{equation}
which allows the tracer displacement to be expressed in terms of the density field as}
\begin{equation}\label{Y1-equn-SM}
Y_1 = \frac{1}{\rhobar} \int_0^{\infty} dx \,[q_1(x,T) - q_1(x,0)] \, \,.
\end{equation}

Also, substituting \eqref{pert1} and \eqref{pert2} in \eqref{optimal-eqns-SM}, we obtain {the Euler-Lagrange equations} to first order in $\lambda$ :
\begin{align}
\partial_{t} q_1(x,t) - D(\rhobar) \partial_{xx} q_1(x,t) & =  -\sigma(\rho)\, \partial_{xx} p_1(x,t) \, \, . \label{q1-evol-SM}
\\
\partial_{t} p_1(x,t) + D(\rhobar) \partial_{xx} p_1(x,t) & =  0 \label{p1-evol-SM} 
\end{align}
{The
boundary condition \eqref{finp2-SM} at leading order in $\lambda$ becomes 
\begin{equation}
p_1(x,T) = \Theta(x)/\rhobar \label{p1-bc-genf-SM} 
\end{equation}
The equations for $p_1$ are now closed, and can be solved as}
\begin{equation}\label{p1xt-SM}
p_1(x,t) = \frac{1}{2\rhobar} {\rm erfc} \left(\frac{-x}{\sqrt{4 D(\rhobar) (T-t)}} \right) \, .  
\end{equation}

For the boundary condition \eqref{inip-gen-SM}, use \eqref{F-quad-alpha}  to obtain $\delta F/\delta q \approx \lambda q_1/(\rhobar\alpha_{\rm ic})$ and hence
\begin{align}
q_1(x,0) & =  \alpha_{\rm ic} \rhobar [  p_1(x,0) - \Theta(x)/\rhobar  ] 
\nonumber\\
& = \alpha_{\rm ic}\rhobar  [p_1(x,0) - p_1(x,T) ] \label{q1-bc-genf-SM}\, \,.   
\end{align}
where the second equality used \eqref{p1-bc-genf-SM}.
(This is one point where the general initial condition enters, and the computation differs from~\cite{sadhu-prl}.)  

{Hence we have derived the equations for the instanton, corresponding to (\eqInst) of the main text.  As described in the main text, this allows the initial condition of the least-action path to be identified, corresponding to an imbalance of density to the left and right of the tracer.}

\subsection{{Tracer variance for HU initial condition}}

{For the HU initial condition ($\alpha_{\rm ic}=0$), the variance of $X(T)$ can now be computed, again following~\cite{sadhu-prl}.
Plugging \eqref{Y1-equn-SM} and \eqref{q1-bc-genf-SM} into \eqref{varX-F2-SM} and setting $\alpha_{\rm ic}=0$ yields }
%
%
\begin{equation}
\Delta X^2_{\rm noise}(T) = \operatorname{Var}[X(T)] = \frac{2}{\rhobar} \int_0^{\infty} dx\, q_1(x,T) - \sigma(\rhobar) \int_0^{T} dt \int_{-\infty}^{\infty} dx \,(\partial_x p_1(x,t))^2
\end{equation}
{where we have identified $\Delta X^2_{\rm noise}(T)$ with the tracer variance computed at $\alpha_{\rm ic}=0$.}
Expressing $q_1(x,t)$ as a product of forward and backward diffusion propagators, followed by some algebra~\cite{sadhu-prl}, one finds
\begin{equation}\label{uni-mftresult-SM}
\Delta X^2_{\rm noise}(T) = \frac{\sigma(\rhobar)}{\rhobar^2} \sqrt{\frac{T}{2 \pi D(\rhobar)}} \, \, .
\end{equation}
This result is quoted in (\eqtrdyn) of the main text. {This computation shows that any HU initial state ({i.e., including states with short-ranged and finite density correlations}) will lead to \eqref{uni-mftresult-SM}, and that this result is not restricted to the quenched case considered in~\cite{sadhu-prl} (where the initial state has no density fluctuations at all).}
\vspace{0.5 cm}

\subsection{{Tracer variance for general initial condition}}

We now calculate $\operatorname{Var}[X(T)]$ for $\alpha_{\rm ic}>0$. Noting the linearity of \eqref{q1-evol-SM}, the least-action path $q_1(x,t)$ can be divided into two contributions, similar to~\cite{sadhu-prl}:
\begin{equation}
q_1(x,t) = q_{\rm I}(x,t) + q_{\rm h} (x,t),
\end{equation}
where $q_{\rm h}(x,t)$ represents the solution of the homogeneous equation
\begin{equation}\label{qh-evol}
\partial_{t} q_{\rm h}(x,t) - D(\rhobar) \partial_{xx} q_{\rm h}(x,t) = 0,
\end{equation}
with the inhomogeneous boundary condition
\begin{equation}\label{qh-bc}
q_{\rm h}(x,0) =  \alpha_{\rm ic} \rhobar\, [p_1(x,0) - p_1(x,T)] \, \,,
\end{equation}
while $q_{\rm I}$ satisfies the inhomogeneous equation
\begin{equation}\label{qI-evol}
\partial_{t} q_{\rm I}(x,t) - D(\rhobar) \partial_{xx} q_{\rm I}(x,t) = -\sigma(\rhobar)\, \partial_{xx} p_1(x,t) \, \, ,
\end{equation}
with homogeneous boundary condition $q_{\rm I}(x,0)=0$. 

{Comparing with the case $\alpha_{\rm ic}=0$ considered above, observe that $q_{\rm I}(x,t)$ is the same least-action path as we already considered, for the HU initial condition.  Noting that the solution \eqref{p1xt-SM} for $p_1$ is independent of $\alpha_{\rm ic}$, 
and that $q_{\rm I}(x,0)=0$, Eq.~\eqref{varX-F2-SM} becomes} 
\begin{equation}\label{tracer-difference}
\operatorname{Var}[X(T)] - \Delta X^2_{\rm noise}(T)  = \frac{2}{\rhobar} \int_0^{\infty} dx\, [q_{\rm h}(x,T) - q_{\rm h}(x,0)] - \frac{1}{\rhobar \alpha_{\rm ic}} \int_{-\infty}^{\infty} dx \, (q_{\rm h}(x,0))^2
\end{equation}

{It only remains to compute the right hand side of \eqref{tracer-difference}, as in~\cite{sadhu-prl}.
To obtain a preliminary result, combine \eqref{p1-evol-SM} and \eqref{qh-evol} to obtain
\begin{equation}
\partial_t[p_1(x,t)q_{\rm h}(x,t)] = - D(\rhobar) \, \partial_x [q_{\rm h}(x,t)  \partial_x p_1(x,t) - p_1(x,t) \partial_x q_{\rm h}(x,t) ].
\end{equation}
This is a continuity equation for $p_1 q_{\rm h}$, so the integral of that quantity is conserved:
\begin{equation}
\frac{\partial}{\partial t} \int_{-\infty}^\infty \!dx\, p_1(x,t) q_{\rm h}(x,t) = 0 \; .
\label{equ:cont}
\end{equation}
Now 
write $I = \int_{0}^{\infty} dx \, [q_{\rm h}(x,T) - q_{\rm h}(x,0)]/\rhobar$ for the first term on the right hand side of \eqref{tracer-difference}.  Then
\begin{eqnarray}\label{identity}
I &=& \int_{-\infty}^{\infty} dx \, \frac{\Theta(x)}{\rhobar} [q_{\rm h}(x,T) - q_{\rm h}(x,0)] 
\nonumber
\\ 
 &=& \int_{-\infty}^{\infty} dx \, p_1(x,T)q_{\rm h}(x,T) -  \int_{-\infty}^{\infty} \! dx \,p_1(x,T)  q_{\rm h}(x,0)  
 \nonumber 
 \\ 
 &=& \int_{-\infty}^{\infty} dx \, p_1(x,0)q_{\rm h}(x,0) -  \int_{-\infty}^{\infty} \! dx \,p_1(x,T) \, q_{\rm h}(x,0) \, ,
\end{eqnarray}
where the second step uses \eqref{p1-bc-genf-SM}, and the final step uses \eqref{equ:cont}.
Combining the integrals and using \eqref{qh-bc} to substitute for $p_1(x,T)-p_1(x,0)$,  we obtain
\begin{equation}\label{ident2}
I = \frac{1}{\rhobar \alpha_{\rm ic}} \int_{-\infty}^{\infty} dx \, (q_{\rm h}(x,0))^2 \,.
\end{equation}
Hence the right hand side of  \eqref{tracer-difference} can be simplified as $2I-I=I$, and using (\ref{qh-bc},\ref{p1xt-SM}) yields
\begin{equation}
\operatorname{Var}[X(T)]- \Delta X^2_{\rm noise}(T)= \alpha_{\rm ic} \rhobar   \int_{-\infty}^{\infty} dx \left[ \frac{\Theta(x)}{\rhobar} - \frac{1}{2\rhobar}{\rm erfc}\left(\frac{-x}{\sqrt{4 D(\rhobar) T}} \right)\right]^2  \\
 \end{equation}
 Performing the integral yields the final result
\begin{equation}\label{main-result-interaction-SM}
\operatorname{Var}[X(T)] =  \Delta X^2_{\rm noise}(T) + \frac{\alpha_{\rm ic}}{ \rhobar} (\sqrt{2}-1)\sqrt{\frac{2 D(\rhobar) T}{\pi}} \, \,.
\end{equation}
from which we identify (see (\eqtrdens) of the main text)
\begin{equation}
\Delta X^2_{\rm dens}(T) = \frac{\sqrt{2}-1}{ \rhobar} \sqrt{\frac{2 D(\rhobar) T}{\pi}}
\end{equation}
and hence
\begin{equation}
\operatorname{Var}[X(T)] = \Delta X^2_{\rm noise}(T) + \alpha_{\rm ic}\,  \Delta X^2_{\rm dens}(T) \,,
\end{equation}
as given in (\eqtrvar) of the main text.
}